\documentclass[journal]{IEEEtran}
\usepackage{amsmath}
\usepackage[noadjust]{cite}
\usepackage{graphicx}
\usepackage{array}
\usepackage{booktabs}
\usepackage{stfloats}
\usepackage{url}
\providecommand{\captionsetup}[1]{}

\begin{document}

\title{A Structured Framework for Calibrating Stochastic Car-Following Models: Data Adequacy, Parameter Sensitivity, and Objective Selection}

\author{
        Shirui Zhou, \IEEEmembership{Student~Member,~IEEE,}
        Junzhe Ding,
        Junfang Tian$^{*}$,
        Shiteng Zheng,
        Rui Jiang,
        Anci Shi

        \thanks{This work is supported by the National Natural Science Foundation of China (Grant No. 72222021, W2411064, 72431006, 72401022)  and Beijing-Tianjin-Hebei Natural Science Foundation Cooperation Project (No. G2024210009).}
        \thanks{$^{*}$Corresponding author.}
        \thanks{Shirui Zhou, Junzhe Ding, Junfang Tian and Anci Shi are with the Institute of Systems Engineering, College of Management and Economics, Tianjin University, Tianjin 300072, China (e-mail: jftian@tju.edu.cn).}
        \thanks{Shiteng Zheng and Rui Jiang are with the School of Systems Science, Beijing Jiaotong University, Beijing 100044, China.}
        }

\markboth{}{Zhou \MakeLowercase{\textit{et al.}}: Calibrating Stochastic Car-Following Models}

\maketitle

\begin{abstract}
Calibrating a stochastic car-following model is harder than calibrating its deterministic counterpart: the loss itself becomes a random variable, so a favorable random realization can be mistaken for a good parameter vector, and it depends on which trajectories are sampled, which variables enter it, and how many realizations are evaluated. This difficulty has left stochastic calibration far less studied than deterministic calibration, and practitioners often default to deterministic-era guidance -- on parameter sensitivity, noise handling, and spacing-versus-speed calibration -- without re-examining whether it survives. This paper develops a structured framework for calibrating stochastic car-following models: a completeness-controlled synthetic design, a corrected and convergence-checked variance-based sensitivity analysis (VBSA), and the minimum-realization (MRMIN) calibration protocol applied at NGSIM scale across two structurally different stochastic mechanisms, QIDM and IDM2D. Within this framework, we test two claims carried over from deterministic calibration -- that a small number of parameters, and the trajectory itself above all, dominates the sensitivity ranking, and that spacing calibration, proven both theoretically and numerically to outperform speed calibration in deterministic models, keeps that guarantee once dynamics are stochastic -- and ask a further question specific to stochastic extensions: whether a model's noise term can be calibrated on its own, without recalibrating its deterministic parameters. A balanced synthetic experiment crosses ten trajectory pairs with eight levels of driving-regime completeness. Variance-based sensitivity analysis doubles its base sample size from 128 to 1024 across three independent designs. At $N=1024$, completeness has a mean total effect of 0.611 for spacing error and 0.639 for speed error, on par with the desired-speed parameter (0.735 and 0.580) and both roughly two orders of magnitude above the residual effect of pair identity (0.006) once regime coverage is isolated as its own factor -- refining, not reversing, the deterministic finding that trajectory identity alone explains most of the variance. The empirical experiment follows the minimum-realization (MRMIN) protocol directly, calibrating 1644 Next Generation SIMulation (NGSIM) trajectories individually for 29 objectives under two stochastic mechanisms, QIDM and IDM2D. Fixing the deterministic parameters at a single population-wide value and calibrating only the noise parameter more than doubles median spacing error relative to full-parameter calibration, but fitting those deterministic parameters per trajectory first and calibrating the noise parameter on top of them recovers essentially all of that accuracy -- even though the fitted noise parameter itself still saturates at its bound rather than settling on a stable value. Cross-dimensional testing shows that spacing calibration remains more robust than speed calibration on average, but the deterministic guarantee that this dominance can never reverse is violated in 19--26\% of trajectories for both mechanisms. Once dominance is no longer guaranteed, a multi-objective screen in relative-error space favors joint spacing--speed goodness-of-fit functions over single-dimension spacing calibration for both mechanisms. Deterministic calibration guarantees should therefore be re-tested, not assumed, once a model is stochastic.
\end{abstract}%


\IEEEpeerreviewmaketitle

\section{Introduction}

Microscopic car-following models underlie the simulators and controller testbeds used to evaluate signal timing, connected-vehicle, and automated-driving systems, so their value for these intelligent transportation systems (ITS) applications depends on calibration. For a stochastic model, however, the loss changes with the simulated random path: estimates then depend on which trajectories are sampled, which variables enter the loss, and how many paths are evaluated, and these choices must be specified before the result can be trusted.

Car-following models expose this problem directly because microscopic simulation propagates small acceleration differences through speed and spacing~\cite{hoogendoorn2001state}. Deterministic formulations include the Gipps model~\cite{gipps1981behavioural}, the Optimal Velocity Model~\cite{bando1995dynamical}, and the Intelligent Driver Model (IDM)~\cite{treiber2000congested,treiber2013traffic,kesting2010enhanced}. Their calibration has been studied through parameter sensitivity, trajectory design, performance variables, and goodness-of-fit (GoF) functions~\cite{treiber2013microscopic,punzo2016speed,punzo2021calibration,ciuffo2012identifiability,punzo2012can,hoogendoorn2010calibration}.

These deterministic studies show that changing the trajectory sample or error definition can change the fitted parameters even when the optimizer is unchanged~\cite{ciuffo2014sensitivity,punzo2014we,punzo2005analysis,ciuffo2010verification,jie2013calibration,ciuffo2013no,punzo2009parameters}. Two results from this literature are commonly treated as general. First, a small number of parameters dominates the sensitivity ranking regardless of context, with the identity of the trajectory itself mattering more than any single parameter~\cite{punzo2014we}. Second, calibrating on spacing weakly dominates calibrating on speed: because spacing is the time integral of speed, spacing-optimal parameters degrade speed much less than speed-optimal parameters degrade spacing, and in a noise-free model this degradation can never turn negative~\cite{punzo2016speed}. Stochastic formulations add a second source of variation on top of this: the simulator can return different losses for the same parameters, so a favorable random realization can be mistaken for a good parameter vector~\cite{punzo2021calibration,zhou2025calibration,xu2020statistical,kendziorra2016stochastic}.

Stochastic IDM extensions and Langevin-type models represent driver variation and unresolved disturbances explicitly~\cite{tian2016empirical,tran2021transport,jiang2018experimental,tian2021car,brockfeld2004calibration}, yet calibrating them has drawn far less attention than calibrating their deterministic counterparts, because a stochastic simulator's loss is itself a random variable: a single realization cannot separate a genuinely well-fitting parameter vector from a lucky random path, so the calibration problem is harder to pose, let alone solve, than its deterministic version. Zhou et al.~\cite{zhou2025calibration} were the first to place this realization problem at the center of stochastic calibration. They showed that single-realization losses are unreliable rankers of candidate parameters and introduced a minimum-error rule across $K$ repeated realizations (MRMIN) that recovers a stable per-trajectory fit despite that noise, establishing the protocol this paper builds on. Yet whether the two deterministic results above survive once the model itself is stochastic is untested; Zhou et al. did not revisit whether deterministic sensitivity rankings or spacing dominance still hold once MRMIN is applied at scale.

Trajectory completeness affects deterministic parameter estimates: Sharma et al.~\cite{sharma2019more} showed that specific driving regimes drive specific calibration errors in a model-dependent way (e.g., standstill for the IDM, cruising for Newell's model), that no single regime maps one-to-one onto the IDM's acceleration parameters the way the cruising regime maps onto Newell's desired speed, and that completeness matters most clearly in validation, where parameters calibrated from more complete trajectories generalize better~\cite{ciuffo2018capability,wang2025segment}. Sensitivity analysis can identify the parameters that influence a stated output~\cite{saltelli2010variance,punzo2014we,saltelli1995sensitivity}, but whether that identification itself depends on completeness remains open. The following questions connect these results to stochastic calibration:

\begin{enumerate}
    \item Does driving-regime completeness reshape parameter sensitivity in a stochastic model, beyond the error-magnitude effect reported for deterministic models?
    \item Is two-stage calibration -- fitting the deterministic parameters first and then the noise parameter on top of them -- sufficient, or does joint full-parameter calibration remain necessary?
    \item Does the deterministic guarantee that spacing calibration weakly dominates speed calibration survive under stochastic dynamics?
    \item Given (3), how should multiple calibration targets be aggregated into a single ranking of goodness-of-fit definitions?
\end{enumerate}

Motivated by these questions, this study follows the MRMIN protocol of Zhou et al.~\cite{zhou2025calibration} directly, calibrating each trajectory's parameters from the minimum error among $K=100$ repeated realizations, and extends it along five directions: (1) a completeness-controlled synthetic design linking regime coverage to parameter sensitivity; (2) a corrected, convergence-checked implementation of variance-based sensitivity analysis (VBSA); (3) a per-trajectory MRMIN screen of 1644 NGSIM trajectories, two orders of magnitude larger than prior demonstrations of this protocol; (4) replication on IDM2D, whose noise enters through a behavioral parameter rather than additively, to test whether the same deterministic results break down under a structurally different stochastic mechanism; and (5) a direct, large-sample test of the spacing--speed dominance argument of~\cite{punzo2016speed} and of two-stage calibration sufficiency, both under stochastic dynamics.

Section II defines QIDM, the trajectory-completeness design, and the VBSA method; Section III reports the calibration workflow and results; Section IV summarizes the findings and limitations.

\section{Basic Setup and Theoretical Background}

\subsection{Model Selection and Calibration Settings}

The Intelligent Driver Model (IDM), proposed by Treiber et al.~\cite{treiber2000congested}, is a widely adopted car-following model and frequently serves as a benchmark due to its capability to reproduce essential traffic dynamics. It also provides a natural foundation for stochastic extensions. In this study, we employ the QIDM model, a Gaussian white-noise extension of the IDM with stochastic intensity $Q$~\cite{treiber2017intelligent}. The QIDM preserves the deterministic structure of the IDM while introducing stochastic acceleration fluctuations to represent unobservable driver variability. The governing equation is

\begin{equation}\label{equ:4.1}
a_n(t) = a_{\mathrm{max}} \left[1 - \left(\frac{v_n(t)}{v_0}\right)^4 - \left(\frac{s^*_n(t)}{s_n(t)}\right)^2 \right] + \xi_n(t)
\end{equation}

with

\begin{align*}
s^*_n(t) &= s_0 + \max \left( v_n(t) T + \frac{v_n(t) \Delta v_n(t)}{2\sqrt{a_{\mathrm{max}}\, b}},\ 0 \right), \\
\Delta v_n(t) &= v_n(t) - v_{n+1}(t), \\
\xi_n(t) &\sim \mathcal{N}(0, Q).
\end{align*}

Here, $a_n(t)$ denotes follower acceleration, $v_n(t)$ follower speed, $v_{n+1}(t)$ leader speed, and $s_n(t)$ the inter-vehicle gap. The relative speed $\Delta v_n(t)$ is positive when the follower is faster. The parameters $a_{\mathrm{max}}$, $b$, $s_0$, $T$, and $v_0$ retain their standard IDM interpretations. In the implemented discrete-time model, $Q$ is acceleration-noise variance per 0.1-s simulation step and $\sqrt{Q}$ is its standard deviation.

\subsubsection*{Calibration setup}

The synthetic sensitivity experiment and the empirical calibration use different parameter bounds. Table~\ref{tab:bounds} states them explicitly. The synthetic generating vector is common to all eight completeness levels. The broad VBSA bounds test influence over plausible mechanisms, whereas the narrower empirical bounds match those used in prior calibration studies on the same data~\cite{punzo2014we,zhou2025calibration}. The VBSA bound for $v_0$ (15--30 m/s, i.e. 54--108 km/h) places the generating value (80 km/h $\approx$ 22.2 m/s) close to the range's center, so the design perturbs desired speed roughly symmetrically around the truth rather than sampling mostly above it. Desired speed is stored in km/h and converted internally to m/s. In the discrete simulator, $Q$ is acceleration-noise variance per time step, with unit $\mathrm{m^2/s^4}$.

\begin{table}[htbp]
\centering
\caption{QIDM parameters and experiment-specific bounds}
\label{tab:bounds}
\scriptsize
\setlength{\tabcolsep}{2.5pt}
\begin{tabular}{lcccc}
\hline\hline
Parameter & Unit & Truth & VBSA bounds & Calibration bounds \\
\hline
$v_0$ & km/h & 80 & [54.0, 108] & [40, 100] \\
$a_{\max}$ & m/s$^2$ & 0.85 & [0.5, 4.5] & [0.5, 3.0] \\
$b$ & m/s$^2$ & 1.50 & [0.5, 4.5] & [0.5, 5.0] \\
$s_0$ & m & 2.00 & [0.5, 10.0] & [0.5, 5.0] \\
$T$ & s & 0.60 & [0.1, 3.0] & [0.1, 1.0] \\
$Q$ & m$^2$/s$^4$ & 0.30 & [0.1, 1.5] & [0, 2.0] \\
\hline\hline
\end{tabular}
\end{table}

\subsection{Data Preparation Considering Trajectory Completeness}

The dataset consists of both empirical and synthetic paired vehicle trajectories. The empirical data are drawn from the NGSIM I80-1 dataset, containing 1649 car-following trajectory samples. Raw NGSIM trajectories carry known tracking error; the series here were reconstructed beforehand with the Kalman filter of Punzo et al.~\cite{punzo2005nonstationary}, so reported errors reflect model--data mismatch, not sensor noise.

To examine parameter recoverability under controlled conditions, we generate eight levels of trajectory completeness following Sharma et al.~\cite{sharma2019more}: ADF, ADFS, CADF, CADFS, FaADF, FaADFS, FaCADF, and FaCADFS. A, D, F, S, C, and Fa denote acceleration, deceleration, following, stopping, cruising, and free acceleration. Each level contains ten trajectories. The leader paths and initial conditions follow the original controlled design, while every follower path is regenerated with the common truth in Table~\ref{tab:bounds}. Independent deterministic noise streams are assigned to the 80 completeness--pair cells. This balanced design varies regime coverage without changing the generating parameters.

Although prior studies have acknowledged that missing driving regimes can distort calibration outcomes~\cite{khattak2017analysis,sharma2018pattern}, trajectory completeness has not been systematically incorporated into stochastic calibration analysis. Importantly, higher completeness does not automatically guarantee lower calibration error; rather, it ensures that sufficient behavioral information is available for parameter identification. In this study, trajectory completeness is treated as a primary indicator of data adequacy.

\subsection{Variance-Based Sensitivity Analysis}

To quantify parameter influence and interaction effects in stochastic car-following models, VBSA is adopted. It decomposes total output variance into contributions from main effects and interaction effects based on Sobol' variance decomposition.

For a model
\[
Y = f(X_1, X_2, \dots, X_k),
\]
the total variance can be decomposed as
\[
V(Y) = \sum_{i=1}^k V_i + \sum_{i<j} V_{ij} + \dots + V_{12\cdots k},
\]
where $V_i$ represents the main effect of $X_i$, and $V_{ij}$ captures pairwise interactions. Higher-order terms follow analogously.

Two sensitivity indices are computed:

\begin{itemize}
\item First-order index:
\[
S_i = \frac{V_{X_i}\!\left[E_{X_{\sim i}}(Y \mid X_i)\right]}{V(Y)}.
\]
\item Total-effect index:
\[
ST_i = 1 - \frac{V_{X_{\sim i}}\!\left[E_{X_i}(Y \mid X_{\sim i})\right]}{V(Y)}.
\]
\end{itemize}

The difference $ST_i - S_i$ reflects interaction effects.

\subsubsection*{Numerical implementation}

A Monte Carlo framework is employed. Two independent Latin hypercube matrices are recombined using Saltelli's design: we call one complete instance of this base-sample-plus-recombination construction a design, so that repeating it three times over independently drawn Latin hypercubes yields three independent replicate estimates of every index. PairID and ComID are sampled directly from discrete uniform distributions. Each row of a design specifies both a parameter vector and a path, i.e., a random seed for the stochastic noise process $\xi_n(t)$ that the simulator plays out over the trajectory. Matched $A$, $B$, and $C_i$ rows use the same path, so each factor contrast is not confounded by a different noise realization; two independent paths are then simulated for every row and their outputs averaged, so that a single unlucky realization cannot drive the estimate.

Instead of analyzing raw accelerations, VBSA is conducted on calibration-relevant outputs: the root-mean-square error (RMSE) of speed, $\mathrm{RMSE}(v)$, and of spacing, $\mathrm{RMSE}(s)$. This ensures that sensitivity results directly inform calibration design.

First-order indices are estimated using Saltelli's scheme, and total-effect indices are computed using Jansen's estimator. The pooled analysis is repeated at base sample sizes 128, 256, 512, and 1024 -- successive powers of two, as required by the Sobol' sequence underlying Saltelli's design -- so that convergence can be checked as the sample size doubles. Each size uses three independent sampling designs and two stochastic paths per model evaluation. Convergence is assessed from the change in mean total effects and the Spearman correlation of factor rankings.

The indices screen practical influence for the stated trajectories and error output; they do not prove structural identifiability. In particular, a small effect of the noise parameter on mean RMSE does not imply a small effect on trajectory dispersion. We therefore do not fix parameters from the pooled indices alone.

\section{Calibration Workflow and Results}

This section applies the calibration workflow in four steps. We first assess data adequacy from driving-regime completeness and parameter sensitivity. We then calibrate every trajectory individually under the MRMIN protocol of Zhou et al.~\cite{zhou2025calibration} for both stochastic models. The remaining steps test whether two-stage calibration -- fitting the deterministic parameters first and the noise parameter on top of them -- is sufficient, whether spacing calibration still weakly dominates speed calibration, and, if not, how the 29 objectives rank once evaluated jointly.

\subsection{Data Adequacy as a Precondition for Calibration}
Calibration of stochastic car-following models is fundamentally constrained by the informational content of trajectory data. If key driving mechanisms embedded in the model (e.g., free acceleration, cruising, deceleration, stopping, and steady following) are insufficiently represented, the calibration problem may be weakly constrained and the resulting sensitivity rankings can be misleading. We therefore start by diagnosing data adequacy, with explicit emphasis on trajectory completeness, before proceeding to parameter screening.

\subsubsection{Why trajectory completeness is needed}
Trajectory identity and duration are common data descriptors, but neither states which driving regimes are present. Two trajectories of equal length can place very different constraints on a car-following model if one contains only steady following and the other also contains free acceleration, cruising, and stopping. We therefore represent data adequacy by the driving regimes covered by a trajectory. Each leader--follower trajectory pair also keeps its own identifier, PairID, so that this pair-level effect can be separated from the effect of regime coverage in the sensitivity design below.

\subsubsection{Controlled verification via trajectory completeness (ComID)}
To separate regime coverage from pair-level heterogeneity, we introduce the trajectory completeness level (ComID). Following Sharma et al.~\cite{sharma2019more}, the synthetic experiment contains eight levels: ADF, ADFS, CADF, CADFS, FaADF, FaADFS, FaCADF, and FaCADFS. Each level contains the same ten PairID levels, so PairID and ComID can be varied independently.

The pooled VBSA includes both identifiers and the six QIDM parameters. Table~\ref{tab:1} reports the total effects at the largest base sample, averaged across three independent designs.

\begin{table}[htbp]
\centering
\caption{Pooled total-effect indices at $N=1024$ (mean $\pm$ SD across three designs)}
\small
\setlength{\tabcolsep}{3.5pt}
\begin{tabular}{lcc}
\hline\hline
Factor & $\mathrm{RMSE}(s)$ & $\mathrm{RMSE}(v)$ \\
\hline
$v_0$ & $0.735\pm0.028$ & $0.580\pm0.013$ \\
$a_{\max}$ & $0.163\pm0.010$ & $0.156\pm0.011$ \\
$b$ & $0.0012\pm0.0001$ & $0.0044\pm0.0003$ \\
$s_0$ & $0.0012\pm0.0001$ & $0.0002\pm0.0000$ \\
$T$ & $0.021\pm0.002$ & $0.007\pm0.001$ \\
$Q$ & $0.0002\pm0.0001$ & $0.0003\pm0.0001$ \\
PairID & $0.0063\pm0.0005$ & $0.0055\pm0.0004$ \\
ComID & $0.611\pm0.030$ & $0.639\pm0.036$ \\
\hline\hline
\end{tabular}
\label{tab:1}
\end{table}

$v_0$ has the largest total effect for spacing RMSE (0.735), narrowly ahead of ComID (0.611); for speed RMSE the order reverses, with ComID largest (0.639) and $v_0$ close behind (0.580). $a_{\max}$ is next for both outputs (0.163/0.156), whereas PairID remains below 0.007. This result isolates regime coverage from pair identity in the balanced design: whether measured against the desired-speed parameter or against the eight-level regime-coverage factor, the residual kinematic variation between otherwise-identical pairs (PairID) is roughly two orders of magnitude smaller than either. It does not imply that driver heterogeneity is negligible in the empirical population.

\begin{figure*}[htbp]
    \centering
    \includegraphics[width=0.94\linewidth]{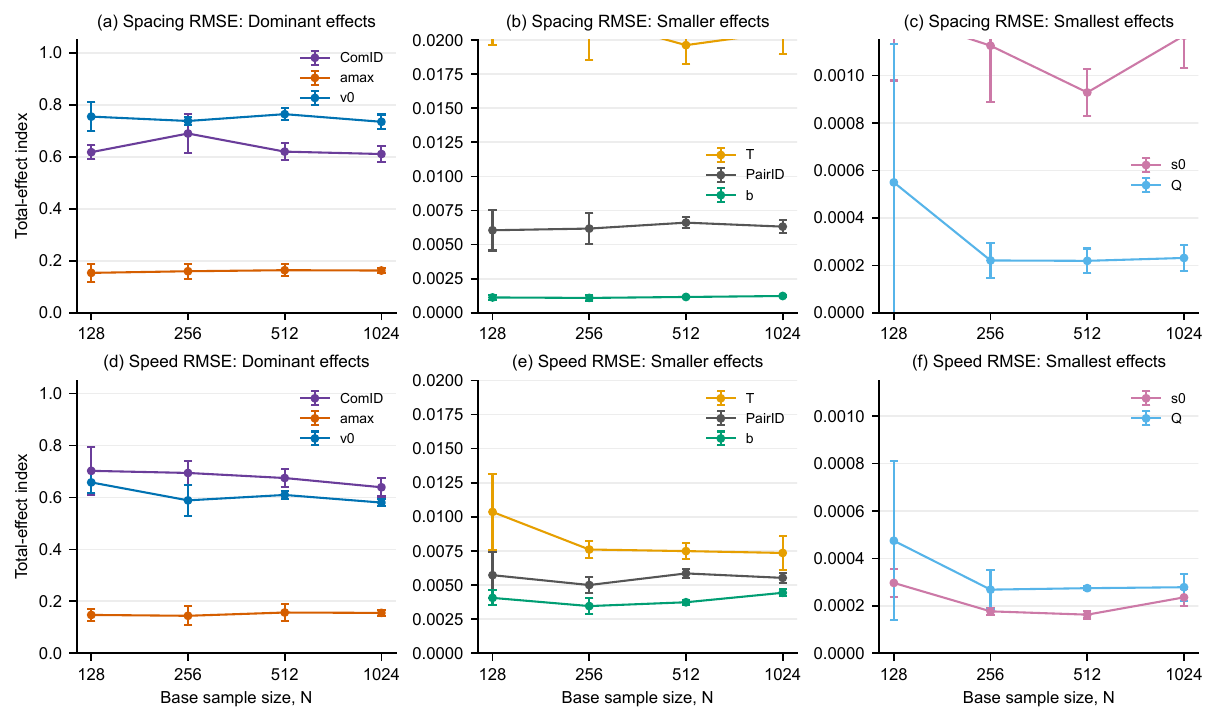}
    \caption{Convergence of QIDM total-effect indices. Points are means across three independent sampling designs; bars are one standard deviation. Each model evaluation averages two matched stochastic paths. The factors are separated into three vertical scales for each error output so that the small effects of $s_0$ and $Q$ remain visible.}
    \label{fig:convergence}
\end{figure*}

Figure~\ref{fig:convergence} shows the sample-size sequence. From $N=512$ to 1024, the largest change in the mean total effect is 0.029 for spacing and 0.036 for speed. The mean absolute changes are 0.0052 and 0.0085. Adjacent factor rankings have Spearman correlations of 0.976--1.000. The ordering of the main factors is therefore stable, although near-zero effects should not be interpreted precisely.

\begin{figure*}[htbp]
    \centering
    \includegraphics[width=0.92\linewidth]{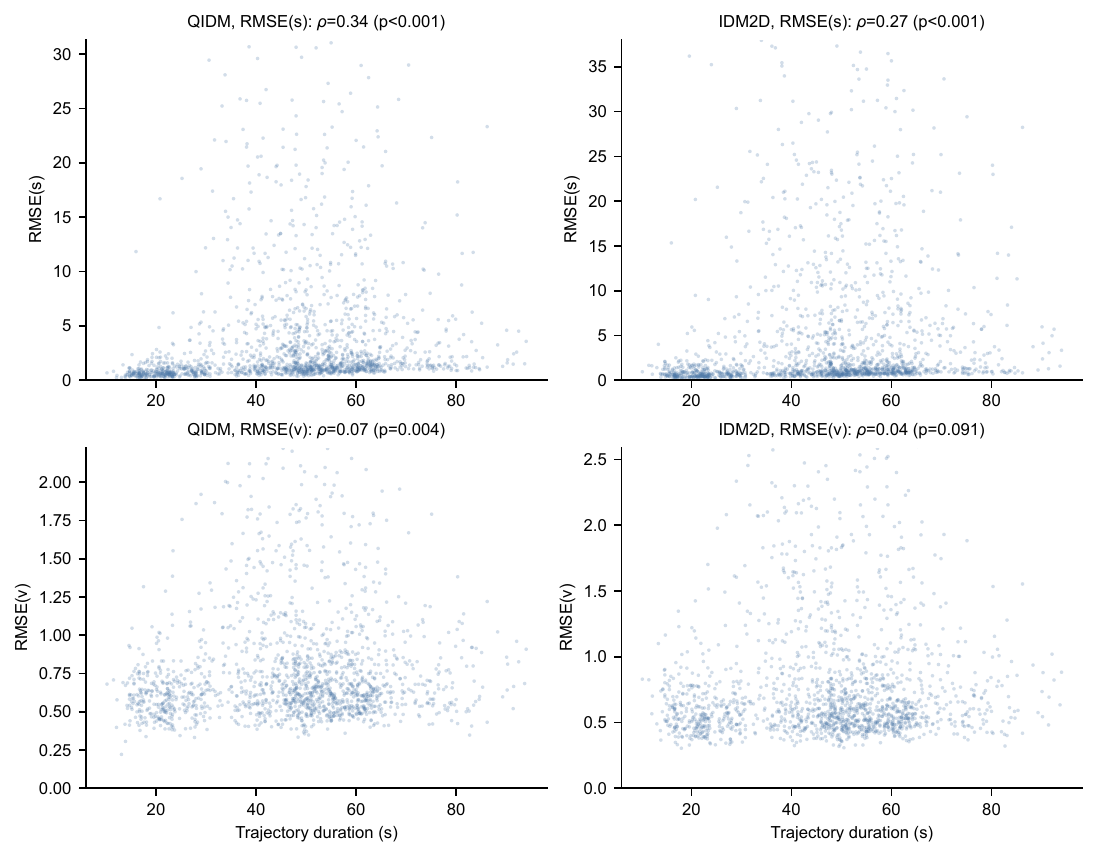}
    \caption{Trajectory duration and MRMIN fit error. Each point is one of 1644 NGSIM trajectories, individually calibrated. QIDM and IDM2D use the corresponding RMSE$(s)$ fit in the spacing panels and RMSE$(v)$ fit in the speed panels. $\rho$ is Spearman's rank correlation, computed on the complete data; the vertical axis of each panel is trimmed to its 98th percentile for legibility.}
    \label{fig:duration}
\end{figure*}

Using the MRMIN fits from Section~III-B, we also test the simpler assumption that a longer trajectory is automatically more informative. Figure~\ref{fig:duration} relates trajectory duration to each trajectory's own fitted RMSE$(s)$ and RMSE$(v)$ across all 1644 trajectories (10.2--94.0~s). Unlike the pooled protocol, per-trajectory MRMIN spacing error rises significantly with duration ($\rho=0.34$, $p<10^{-44}$ for QIDM; $\rho=0.27$, $p<10^{-27}$ for IDM2D), consistent with noise accumulating over a longer horizon even under the best of $K=100$ realizations; speed error shows a much weaker relationship ($\rho=0.07$ for QIDM, $p=0.004$; $\rho=0.04$ for IDM2D, not significant). Duration's association with error therefore depends on the calibration protocol itself and does not by itself indicate which driving regimes a trajectory contains, reinforcing completeness rather than length as the data-adequacy measure.

\subsubsection{Sensitivity rankings depend on completeness level}
\begin{figure*}[htbp]
    \centering
    \includegraphics[width=0.96\linewidth]{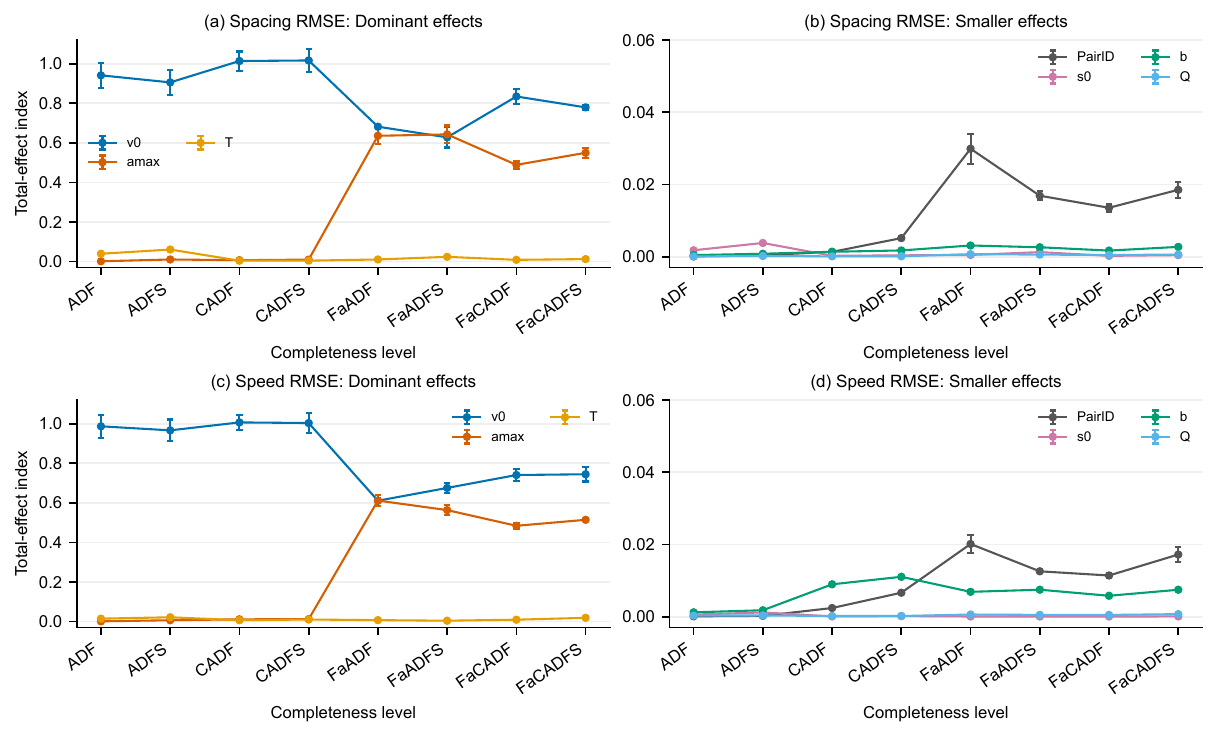}
    \caption{QIDM total-effect indices conditional on trajectory completeness. Points are means across three independent designs at $N=256$; bars are one standard deviation. Each evaluation averages two matched stochastic paths. Dominant and smaller effects use separate vertical scales.}
    \label{fig:conditional-vbsa}
\end{figure*}

We next recompute $S_i$ and $ST_i$ within each completeness level. Each conditional analysis uses a base sample of 256, three independent designs, and two matched stochastic paths per evaluation. Figure~\ref{fig:conditional-vbsa} shows all total effects. Table~\ref{tab:2} reports the two largest spacing-error effects.

\begin{table}[htbp]
\centering
\caption{Dominant spacing-RMSE total effects by completeness level}
\scriptsize
\setlength{\tabcolsep}{1.8pt}
\begin{tabular}{lcccc}
\hline\hline
ComID & Dominant factor & $S_{T_i}$ & Secondary factor & $S_{T_i}$ \\
\hline
ADF     & $v_0$        & $0.941\pm0.062$ & $T$         & $0.040\pm0.003$ \\
ADFS    & $v_0$        & $0.906\pm0.065$ & $T$         & $0.061\pm0.005$ \\
CADF    & $v_0$        & $1.013\pm0.048$ & $a_{\max}$  & $0.007\pm0.000$ \\
CADFS   & $v_0$        & $1.016\pm0.060$ & $a_{\max}$  & $0.010\pm0.000$ \\
FaADF   & $v_0$        & $0.682\pm0.011$ & $a_{\max}$  & $0.636\pm0.043$ \\
FaADFS  & $a_{\max}$   & $0.643\pm0.045$ & $v_0$       & $0.628\pm0.051$ \\
FaCADF  & $v_0$        & $0.834\pm0.040$ & $a_{\max}$  & $0.489\pm0.019$ \\
FaCADFS & $v_0$        & $0.779\pm0.012$ & $a_{\max}$  & $0.550\pm0.024$ \\
\hline\hline
\end{tabular}
\label{tab:2}
\end{table}

Once $v_0$'s bound is centered on the generating value, it leads the spacing-RMSE ranking in seven of the eight completeness levels and the speed-RMSE ranking in all eight (the two totals slightly exceeding 1 for CADF and CADFS reflect Monte Carlo noise in Jansen's estimator when the true index is close to unity, not an inadmissible value). $a_{\max}$ overtakes it only in FaADFS for spacing (0.643 vs.\ 0.628) and ties with it in FaADF for speed (0.611 each), and in both cases the crossover appears only once free-acceleration segments are present in the trajectory. $T$ is a distant secondary factor confined to the two least complete, following-only levels (ADF, ADFS) and never leads. PairID, $b$, $s_0$, and $Q$ remain small under these mean-error outputs. Regime coverage therefore changes less which parameter is identifiable than how closely $a_{\max}$ can rival $v_0$: the desired speed sets an asymptotic target that every regime constrains to some degree, while the acceleration cap becomes comparably informative only once the trajectory contains the free-acceleration maneuvers that exercise it. These results show that a low sensitivity can mean that the relevant driving regime is absent; it is not evidence that the parameter is generally unimportant.

The result differs from earlier deterministic IDM studies. Punzo et al.~\cite{punzo2014we} found that PairID, not any parameter, explains most of the variance (80.4\%/93.3\% for RMSE$(s)$/RMSE$(v)$); among parameters alone, $T$ ranks highest, at a much smaller share (45\%/28\%). ComID, however, is a synthetic-experiment construct with no counterpart in the real, heterogeneous NGSIM trajectories Punzo et al. analyzed, and their PairID also captures genuine driver-to-driver heterogeneity, which the present design holds fixed across pairs (Section~II-B) specifically to isolate regime coverage as an independent factor. Driver heterogeneity therefore remains a real and important source of variation, as Punzo et al. correctly found; the result here is narrower but still substantive: once that heterogeneity is held constant, explicit regime coverage (0.611/0.639) and the desired-speed parameter (0.735/0.580) both have a far larger effect than the residual kinematic variation between pairs (0.006/0.006). Sharma et al.~\cite{sharma2019more} did not test ranking stability across completeness; they found instead that a missing regime degrades several parameters through interaction terms, not only its nominal counterpart, and that the critical regime is model-specific. The conditional QIDM ranking here is consistent with that model-specificity at one remove: $v_0$ leads almost regardless of which regimes are present, but $a_{\max}$'s standing relative to it is specific to the free-acceleration regime. Because the conditional experiment has a smaller sample, this comparison is evidence of regime dependence rather than a claim that stochasticity itself causes the difference.

\subsubsection{Completeness and parameter recovery}
Sensitivity indicates which parameters affect an error output, but it does not show whether their true values can be recovered. We therefore calibrate QIDM separately at each completeness level. For each level, eight of the ten synthetic trajectories are used for calibration and two are held out. RMSE$(s,v)$ and Theil's $U(s,v)$ are each optimized with three seeds of a genetic algorithm (GA). Parameter error is the root mean square difference from the known generating vector after each parameter is divided by its calibration range.

\begin{figure}[htbp]
    \centering
    \includegraphics[width=0.98\linewidth]{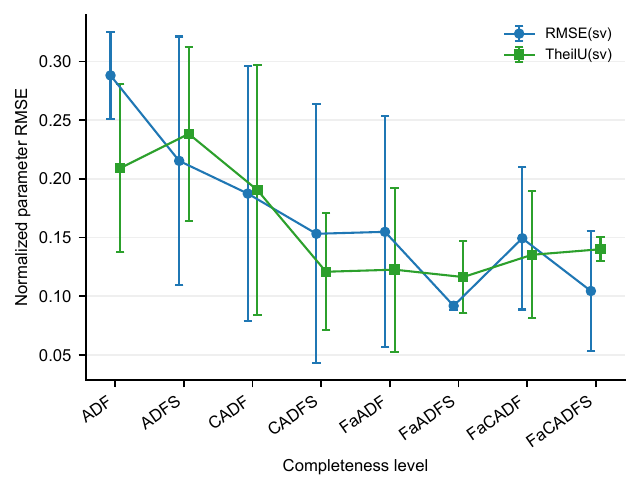}
    \caption{QIDM parameter-recovery error by trajectory completeness. Points are means across three GA seeds; bars are one standard deviation.}
    \label{fig:completeness-calibration}
\end{figure}

Figure~\ref{fig:completeness-calibration} shows a broad reduction in parameter error as more regimes are included. Averaged across the two objectives, the normalized error is 0.249 for ADF and 0.227 for ADFS. It falls to 0.137 for CADFS and to 0.104--0.142 for the four Fa cases. The decline is not strictly monotonic because only ten synthetic trajectories are available per level and the GA remains stochastic. Completeness therefore improves parameter constraint on average; it does not guarantee a lower error for every level, objective, or seed.

\subsection{Objective Definitions and the MRMIN Protocol}

The converged VBSA establishes two requirements for the calibration experiment. First, trajectories must contain enough driving regimes to inform the parameters being estimated. Second, low total effects for $b$, $s_0$, and $Q$ under the present RMSE outputs cannot be treated as proof that these parameters are redundant; sensitivity is conditional on the sampled input ranges, trajectories, and output measures.

We compare candidate objectives under the MRMIN protocol of Zhou et al.~\cite{zhou2025calibration}: each trajectory is calibrated independently, and the loss for a candidate parameter vector is the minimum GoF value among $K=100$ repeated stochastic realizations, following the original protocol directly rather than a pooled alternative.

Let $y_t$ and $\hat y_t$ be observed and simulated values and $e_t=y_t-\hat y_t$. Table~\ref{tab:gof-corrected} defines the seven single-variable functions used throughout the experiment: RMSE; the mean absolute error (MAE); the normalized RMSE (NRMSE); the mean absolute percentage error (MAPE); the root-mean-square percentage error (RMSPE); the integrated RMSPE (IRMSPE); and Theil's $U$. MAE and MAPE have no outer square root. IRMSPE has no extra division by the sample count. NRMSE is normalized by the observed root-mean-square (RMS) value. No final samples are removed. MAPE, RMSPE, and IRMSPE omit observations with $|y_t|\le10^{-6}$ to avoid division by zero.

\begin{table}[htbp]
\centering
\caption{Corrected single-variable GoF definitions}
\label{tab:gof-corrected}
\scriptsize
\setlength{\tabcolsep}{3.3pt}
\begin{tabular}{lc}
\hline\hline
GoF & Definition \\
\hline
RMSE & $\sqrt{N^{-1}\sum_t e_t^2}$ \\
MAE & $N^{-1}\sum_t |e_t|$ \\
NRMSE & $\sqrt{\sum_t e_t^2/\sum_t y_t^2}$ \\
MAPE & $N^{-1}\sum_t |e_t/y_t|$ \\
RMSPE & $\sqrt{N^{-1}\sum_t(e_t/y_t)^2}$ \\
IRMSPE & $\sqrt{(\sum_t e_t^2/|y_t|)/\sum_t|y_t|}$ \\
Theil's $U$ & $\mathrm{RMSE}(e)/[\mathrm{RMS}(y)+\mathrm{RMS}(\hat y)]$ \\
\hline\hline
\end{tabular}
\end{table}

Following the calibration guidelines of Punzo et al.~\cite{punzo2021calibration}, these seven GoF functions are combined with selectable variable sets to give 29 objectives in total: RMSE and MAE are each defined only for the single variables $s$, $v$, and $a$ (three objectives apiece), IRMSPE only for $s$, $v$, and $sv$ (three objectives), and RMSPE, Theil's $U$, MAPE, and NRMSE for all five variable sets $s$, $v$, $a$, $sv$, and $sva$ (five objectives apiece). The 29 objectives retain the original combinations of spacing ($s$), speed ($v$), acceleration ($a$), $sv$, and $sva$. RMSPE, Theil's $U$, MAPE, and NRMSE are already invariant to a common scale factor, which is why only they are extended to the joint variable sets $sv$ and $sva$. Selected variable losses are averaged with equal weight. This avoids directly adding errors expressed in metres, metres per second, and metres per second squared. Observed acceleration is the numerical gradient of measured speed at the 0.1-s sampling interval.

For objective $c$, trajectory $j$, selected variable set $\mathcal M_c$, and stochastic realization $r\in\{1,\dots,K\}$, calibration minimizes over the trajectory's own parameter vector $\beta_j$:
\begin{equation}\label{eq:mrmin-loss}
L_{c,j}(\beta_j)=\min_{r=1,\dots,K}\ \frac{1}{|\mathcal M_c|}\sum_{m\in\mathcal M_c}
g_c(y_{m,j},\hat y_{m,j,r}(\beta_j)).
\end{equation}
Each trajectory therefore receives its own fitted parameter vector, one GA seed, and $K=100$ realizations per candidate.

The empirical data contain 1649 NGSIM I-80 leader--follower trajectories; 1644 have at least 100 samples and are retained. Every retained trajectory is calibrated for all 29 objectives and both models (QIDM and IDM2D), a run using a 24-member Latin-hypercube initial population and 24 generations with tournament selection, bounded arithmetic crossover, Gaussian mutation, and two elites (600 candidate evaluations, each requiring $K=100$ realizations). This gives 95352 independent MRMIN fits, roughly two orders of magnitude more trajectories than prior demonstrations of this protocol.

\subsection{Replication with a Second Stochastic Model}

The QIDM uses additive Gaussian acceleration noise. IDM2D~\cite{tian2016improved} instead perturbs a behavioral parameter, retaining the deterministic acceleration in (\ref{equ:4.1}) while randomly switching the desired time headway; repeating the protocol on this structurally different mechanism tests whether the results below hold beyond additive noise. IDM2D updates its desired headway as
\begin{equation}\label{eq:idm2d-long}
T_{k+1}=\begin{cases}
T_1+U_k\Delta T, & U'_k<p\Delta t,\\
T_k, & \text{otherwise},
\end{cases}
\end{equation}
where $U_k$ and $U'_k$ are independent uniform variables. The switching rate is $p$ and $T_k\in[T_1,T_1+\Delta T]$. The calibrated bounds are $T_1\in[0.1,1]$~s, $\Delta T\in[0.01,1]$~s, and $p\in[0,1]$~s$^{-1}$, together with the same bounds for $v_0$, $a_{\max}$, $b$, and $s_0$ used in the empirical QIDM experiment. The switching-rate parameter $p$ is calibrated together with the other six IDM2D parameters here, correcting an earlier protocol that held it fixed even though the simulator already treated it as free.

\subsection{Is Two-Stage Calibration Sufficient?}

A stochastic model could in principle be simplified by calibrating its deterministic and noise parameters separately rather than jointly. We test two versions of this for QIDM under MRMIN, both against the RMSE$(s)$ objective. The naive version fixes the five non-noise parameters at a single population-wide value -- the median of their full six-parameter RMSE$(s)$ fits across the 1644 trajectories ($v_0=56.98$~km/h, $a_{\max}=0.622$~m/s$^2$, $b=1.079$~m/s$^2$, $s_0=4.257$~m, $T=0.904$~s) -- and calibrates only $Q$ against it. The two-stage version instead fits the five deterministic parameters \emph{per trajectory} first, with $Q$ fixed at 0 (the model is then exactly deterministic, so a single simulation replaces the $K=100$ realizations used elsewhere), and only then fixes that trajectory's own deterministic fit and calibrates $Q$ against it under the full MRMIN protocol.

\begin{figure}[htbp]
    \centering
    \includegraphics[width=0.98\linewidth]{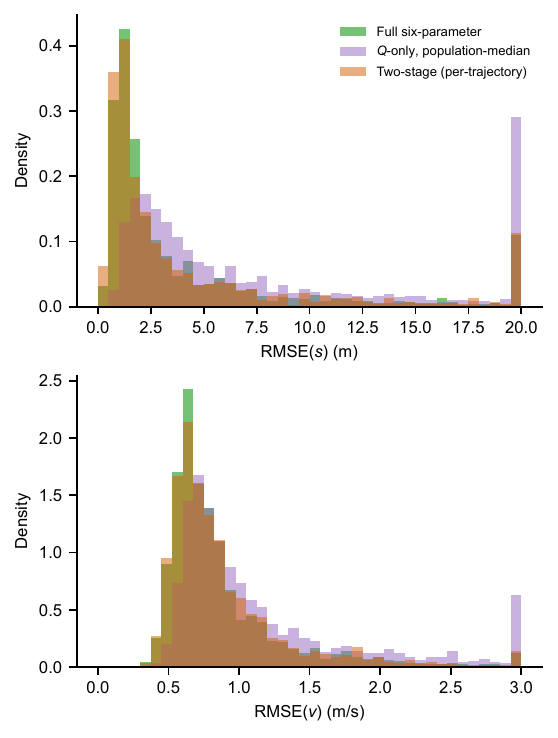}
    \caption{QIDM spacing and speed error distributions across 1644 trajectories: full six-parameter joint MRMIN fits, the population-median $Q$-only fits, and the two-stage (per-trajectory deterministic fit, then $Q$) fits. Both axes are truncated at the plotted limit, and the rightmost bin pools all larger values: RMSE$(s)$ exceeds 20~m for 5.2\% of full-parameter, 14.1\% of population-median $Q$-only, and 5.4\% of two-stage fits (maxima 115~m, 122~m, and 114~m), and RMSE$(v)$ exceeds 3~m/s for 0.7\%, 4.4\%, and 0.9\%, respectively (maxima 5.1~m/s, 8.4~m/s, and 5.7~m/s).}
    \label{fig:qonly}
\end{figure}

Figure~\ref{fig:qonly} compares the three. The naive, population-median version replicates the earlier finding: median RMSE$(s)$ rises from 1.93~m to 4.68~m, and median RMSE$(v)$ rises from 0.74~m/s to 0.91~m/s. But the two-stage version, which differs only in fitting the deterministic stage per trajectory rather than at a single population value, essentially matches the full joint fit: median RMSE$(s)$ is 1.89~m (against 1.93~m for the joint fit) and median RMSE$(v)$ is 0.75~m/s (against 0.74~m/s), with 77\% of trajectories within 10\% of their own joint-fit spacing error. The failure of the naive version is therefore a failure of using a population-level deterministic fit, not evidence that sequential calibration itself is inadequate; done per trajectory, two-stage calibration recovers essentially all of the joint fit's accuracy, at a fraction of its search dimensionality (a 1-parameter search for $Q$ instead of a joint 6-parameter one).

This equivalence in fit quality does not, however, extend to the noise parameter's own value. The fitted $Q$ still clusters near its upper bound in the two-stage version (64\% of trajectories above 1.9 of the 2.0 bound, median exactly at the bound), essentially the same saturation pattern as the naive version (mean 1.88 of 2.0). Even when the deterministic stage is fit correctly, $Q$ under MRMIN's minimum-over-$K$-realizations rule is still being pushed to its bound to exploit that minimum rather than converging to a stable, interior estimate of the acceleration-noise magnitude. Two-stage calibration is therefore sufficient for reproducing a trajectory's spacing and speed errors, but the fitted noise parameter should still not be read as a physically meaningful noise estimate.

\subsection{Does Stochasticity Break Spacing--Speed Dominance?}

Punzo and Montanino~\cite{punzo2016speed} proved that, because spacing is the time integral of speed, calibrating a deterministic car-following model on spacing weakly dominates calibrating on speed: the relative degradation from using the other dimension's parameters is bounded below by zero. Their empirical demonstration, on all 2037 trajectories of an NGSIM I80-1 dataset reconstructed with the same method (Section~II-B) and the deterministic IDM, found mean relative degradation of 63\% for spacing under speed-optimal parameters versus 9\% for speed under spacing-optimal parameters, with both distributions bounded at exactly 0\%. Our own processed set retains 1649 of these trajectories before the 100-sample length filter, so the two counts are not identical, but both draw on the same reconstructed source and reconstruction method.

We repeat this cross-dimensional test under MRMIN for both stochastic models, using the already-fitted RMSE$(s)$- and RMSE$(v)$-optimal parameters $\beta_s,\beta_v$ for each of the 1644 trajectories:
\begin{align}
RE_s&=\frac{\mathrm{RMSE}(s,\beta_v)-\mathrm{RMSE}(s,\beta_s)}{\mathrm{RMSE}(s,\beta_s)},\label{eq:cross-dim-s}\\
RE_v&=\frac{\mathrm{RMSE}(v,\beta_s)-\mathrm{RMSE}(v,\beta_v)}{\mathrm{RMSE}(v,\beta_v)}.\label{eq:cross-dim-v}
\end{align}

\begin{figure}[htbp]
    \centering
    \includegraphics[width=0.98\linewidth]{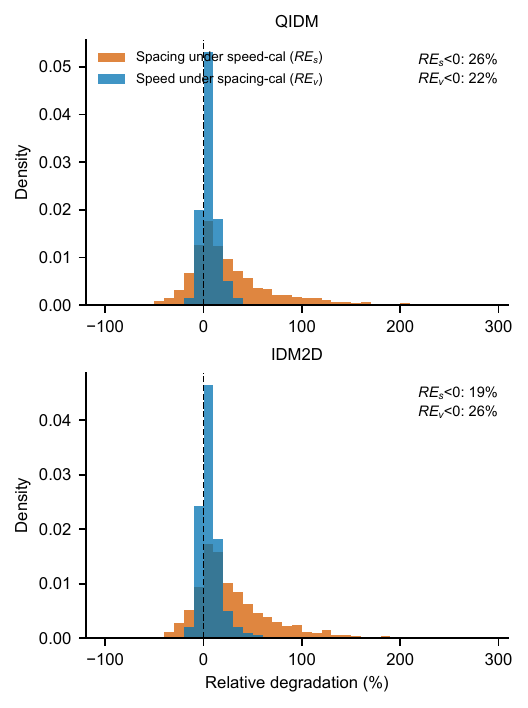}
    \caption{Cross-dimensional relative degradation (\ref{eq:cross-dim-s})--(\ref{eq:cross-dim-v}) across 1644 trajectories for QIDM and IDM2D under MRMIN. The dashed line marks zero, the deterministic floor of~\cite{punzo2016speed}.}
    \label{fig:crossdim}
\end{figure}

Figure~\ref{fig:crossdim} shows both distributions. The overall direction survives: spacing calibration remains more cross-dimensionally robust on average ($RE_v$ mean 6.2\%/6.4\% for QIDM/IDM2D) than speed calibration ($RE_s$ mean 28.8\%/30.8\%). But the deterministic zero floor does not survive: $RE_s$ is negative for 25.6\% of QIDM trajectories and 19.0\% of IDM2D trajectories, and $RE_v$ is negative for 21.7\%/26.4\%, respectively. In roughly one in five to one in four trajectories, calibrating on the other dimension's stochastic realizations accidentally outperforms the dimension-optimal fit, which is impossible under Punzo and Montanino's deterministic argument. Stochastic acceleration noise therefore breaks the deterministic dominance guarantee itself, not merely its average magnitude, under both an additive-noise mechanism (QIDM) and a behavioral-switching mechanism (IDM2D). Part of this gap may reflect finite-budget search on a noisier fitness landscape rather than pure evaluation noise at fixed parameters, but both are consequences of stochasticity rather than of a deterministic model; Punzo and Montanino's own optimizer is imperfect yet reports exactly 0\% negative cases across 2037 trajectories, so the contrast is unlikely to be a search artifact alone.

\subsection{Multi-Objective Ranking under MRMIN}

Because dominance is no longer guaranteed, a single spacing-only target cannot be assumed sufficient, and the 29 objectives are compared jointly in relative-error space. For each model, the baselines $z_s^*$ and $z_v^*$ are the mean RMSE$(s)$ and RMSE$(v)$ achieved by the RMSE$(s)$- and RMSE$(v)$-only objectives across all 1644 trajectories, and $RE_s$, $RE_v$ for every other objective are computed against these baselines with the same definitions as (\ref{eq:cross-dim-s})--(\ref{eq:cross-dim-v}), following~\cite{punzo2016speed}. The point $RE_s=RE_v=0$, where an objective would match both single-variable baselines at once, is the utopia point; an objective's summed distance $RE_s+RE_v$ ranks how close it comes.

\begin{figure}[htbp]
    \centering
    \includegraphics[width=0.98\linewidth]{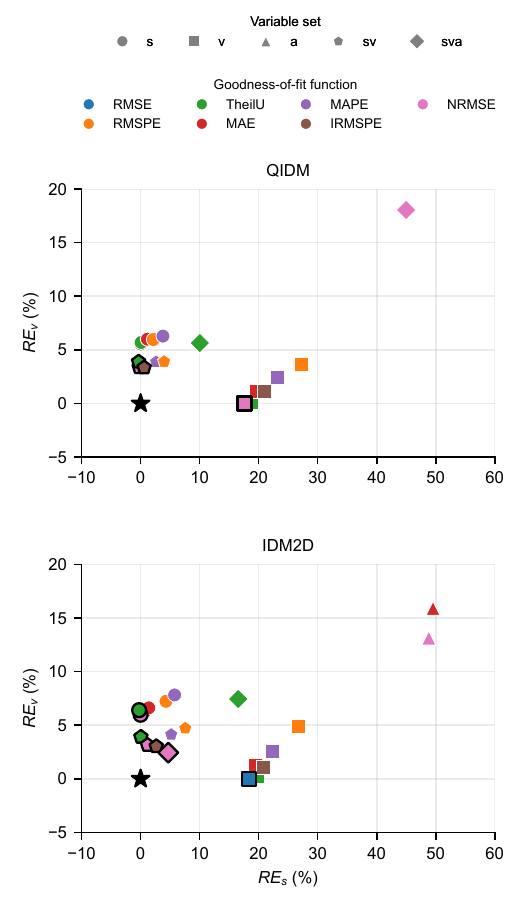}
    \caption{Relative-error space for 29 MRMIN objectives, averaged over 1644 trajectories. The star marks the utopia point; black-edged markers are Pareto efficient. Color identifies the GoF and marker shape identifies the variable set. For a single-variable objective, RMSE and NRMSE share the same minimizer up to a constant scale factor, so their markers coincide exactly (e.g., RMSE$(v)$ and NRMSE$(v)$ for QIDM); such pairs appear as one visible point.}
    \label{fig:utopia}
\end{figure}

Figure~\ref{fig:utopia} shows both models. For QIDM, five objectives are Pareto efficient; the closest to the utopia point is NRMSE$(s,v)$ ($RE_s=-0.2\%$, $RE_v=3.4\%$; the small negative value is within GA search noise around the $z_s^*$ baseline, not a true improvement on it), followed by Theil's $U(s,v)$ and IRMSPE$(s,v)$. For IDM2D, seven objectives are Pareto efficient, led by Theil's $U(s,v)$ ($RE_s=0.0\%$, $RE_v=3.9\%$) and NRMSE$(s,v)$. Four objectives, NRMSE$(s,v)$, Theil's $U(s,v)$, IRMSPE$(s,v)$, and RMSE$(v)$, lie on both fronts. Unlike the single-variable spacing preference reported for the deterministic case, the best MRMIN combinations for both stochastic mechanisms jointly weight spacing and speed with a scale-consistent GoF, one normalized by the observed magnitude (NRMSE, Theil's $U$) rather than expressed in raw physical units (RMSE, MAE). Objectives that add acceleration remain far from the utopia point ($RE_s+RE_v$ exceeding 60\% in every case, and above 125\% for single-variable RMSE$(a)$), consistent with the acceleration instability already noted for percentage-based measures.

\section{Conclusion}

This study developed a structured framework for calibrating stochastic car-following models -- a completeness-controlled synthetic design, a corrected variance-based sensitivity analysis, and the minimum-realization (MRMIN) protocol of Zhou et al.~\cite{zhou2025calibration} applied at 1644-trajectory NGSIM scale across two structurally different stochastic mechanisms -- and used it to test whether stochastic dynamics change two results established for deterministic car-following calibration -- that a small number of parameters, and above all the trajectory itself, dominates the sensitivity ranking, and that spacing calibration weakly dominates speed calibration -- and to ask a further question specific to stochastic extensions: whether a model's noise term can be calibrated on its own, without recalibrating its deterministic parameters. A fourth finding follows directly from the third: once dominance is not guaranteed, the choice among goodness-of-fit definitions must be re-examined too.

First, trajectory content reshapes parameter sensitivity rather than only its magnitude. In the balanced QIDM experiment, completeness has a mean total-effect index of 0.61--0.64 for spacing and speed error, on par with the desired-speed parameter $v_0$ (0.74/0.58) and both far exceeding PairID's 0.006. ComID is a synthetic-experiment construct absent from real trajectories, and the design holds behavioral parameters fixed across pairs specifically to isolate it from driver heterogeneity, which Punzo et al.'s PairID-dominated deterministic finding~\cite{punzo2014we} correctly shows is real and important; the result here is that, once that heterogeneity is held constant, regime coverage still has an effect on par with the model's own most influential parameter and roughly two orders of magnitude larger than the residual kinematic variation PairID captures. Conditionally, $v_0$ leads the ranking in almost every completeness level; only once free-acceleration segments are present does $a_{\max}$ close the gap and occasionally overtake it, and the normalized parameter-recovery error falls from 0.249 (ADF) to 0.104--0.142 (the four Fa cases). Under per-trajectory MRMIN, duration itself correlates with spacing error ($\rho=0.34$/0.27 for QIDM/IDM2D), but this reflects noise accumulating over longer horizons rather than regime coverage, so completeness rather than length should still guide data adequacy.

Second, whether a stochastic model's noise parameter can substitute for full-parameter calibration depends on how the deterministic parameters it is fixed against are chosen. Fixing QIDM's five deterministic parameters at a single population-wide value and calibrating $Q$ alone more than doubles median spacing RMSE (1.93 to 4.68~m); fixing them at each trajectory's own deterministic fit instead and calibrating $Q$ on top of it recovers essentially the same accuracy as joint six-parameter calibration (median spacing RMSE 1.89~m, speed RMSE 0.75~m/s). In both versions, however, the fitted $Q$ is pushed toward its upper bound rather than settling on a genuine noise magnitude, so this two-stage sufficiency is a statement about fit quality, not about the noise parameter's own identifiability.

Third, the deterministic guarantee that spacing calibration weakly dominates speed calibration~\cite{punzo2016speed} does not survive stochastic dynamics. Cross-dimensional relative degradation stays positive on average (28.8--30.8\% for spacing under speed-optimal parameters versus 6.2--6.4\% for speed under spacing-optimal parameters), matching the deterministic direction, but its deterministic zero floor is violated in 19--26\% of trajectories for both QIDM and IDM2D: calibrating on the other dimension's stochastic realization can accidentally outperform the dimension-optimal fit, which the deterministic argument rules out entirely.

Fourth, because dominance is no longer guaranteed, the 29-objective relative-error screen no longer favors single-variable spacing calibration, unlike the deterministic case. For both QIDM and IDM2D, joint spacing--speed objectives with a scale-consistent GoF (NRMSE$(s,v)$, Theil's $U(s,v)$) are closest to the utopia point, with four objectives Pareto efficient for both mechanisms; objectives that add acceleration remain far from it.

The study still uses two related stochastic IDM formulations, one GA family, and one calibration seed per trajectory. NGSIM I-80 is a single freeway site and time window despite its 1644 trajectories. Broader model classes, optimizer families, and data sources are still needed before the specific rankings can be treated as general. The transferable result is that deterministic calibration guarantees should be re-tested, not assumed, once a model's dynamics are stochastic.

\section*{Data and Code Availability}
The NGSIM I-80 source data are publicly available from the U.S. Department of Transportation~\cite{usdotngsim}. The code, corrected metric definitions, random seeds, balanced synthetic-data provenance, and derived result tables are included in the supplementary reproducibility package. No new human-participant data were collected for this study.

\bibliographystyle{IEEEtran}
\renewcommand{\IEEEbibitemsep}{-1.5pt}
\bibliography{IEEEabrv,references}

\end{document}